\documentclass[prx,twocolumn,english,superscriptaddress,floatfix,longbibliography]{revtex4-2} 
\usepackage{graphicx,color}  
\usepackage{dcolumn}   
\usepackage{bm}        
\usepackage{amssymb}   
\usepackage{subfigure}
\usepackage{lipsum}   
\usepackage{physics}

\usepackage[heightadjust=all,valign=t]{floatrow}
\usepackage{fr-subfig}

\usepackage{sidecap}
\usepackage{comment}
\usepackage{mathrsfs}
\usepackage{soul}
\usepackage{comment}
\usepackage{xcolor}
\usepackage{tikz}
\usepackage{amsmath}

\begin{document}

\title{Single Flux Quantum-Based Digital Control of\\ Superconducting Qubits in a Multi-Chip Module}

\author{C. H. Liu} 
\affiliation{Department of Physics, University of Wisconsin-Madison, Madison, Wisconsin 53706, USA}

\author{A. Ballard}
\affiliation{Department of Physics, Syracuse University, Syracuse, New York 13244, USA}

\author{D. Olaya} 
\affiliation{National Institute of Standards and Technology, Boulder, Colorado 80305, USA}
\affiliation{Department of Physics, University of Colorado, Boulder, Colorado 80305, USA}

\author{D. R. Schmidt} 
\affiliation{National Institute of Standards and Technology, Boulder, Colorado 80305, USA}

\author{J. Biesecker} 
\affiliation{National Institute of Standards and Technology, Boulder, Colorado 80305, USA}

\author{T. Lucas} 
\affiliation{National Institute of Standards and Technology, Boulder, Colorado 80305, USA}

\author{J. Ullom} 
\affiliation{National Institute of Standards and Technology, Boulder, Colorado 80305, USA}
\affiliation{Department of Physics, University of Colorado, Boulder, Colorado 80305, USA}

\author{S. Patel} 
\affiliation{Department of Physics, University of Wisconsin-Madison, Madison, Wisconsin 53706, USA}

\author{O. Rafferty} 
\affiliation{Department of Physics, University of Wisconsin-Madison, Madison, Wisconsin 53706, USA}

\author{A. Opremcak} 
\affiliation{Department of Physics, University of Wisconsin-Madison, Madison, Wisconsin 53706, USA}

\author{K. Dodge}
\affiliation{Department of Physics, Syracuse University, Syracuse, New York 13244, USA}

\author{V. Iaia}
\affiliation{Department of Physics, Syracuse University, Syracuse, New York 13244, USA}

\author{T. McBroom}
\affiliation{Department of Physics, Syracuse University, Syracuse, New York 13244, USA}

\author{J. L. DuBois} 
\affiliation{Physics Division, Lawrence Livermore National Laboratory, Livermore, California 94550, USA}

\author{P. F. Hopkins} 
\affiliation{National Institute of Standards and Technology, Boulder, Colorado 80305, USA}

\author{S. P. Benz} 
\affiliation{National Institute of Standards and Technology, Boulder, Colorado 80305, USA}

\author{B. L. T. Plourde}
\affiliation{Department of Physics, Syracuse University, Syracuse, New York 13244, USA}

\author{R. McDermott}
\affiliation{Department of Physics, University of Wisconsin-Madison, Madison, Wisconsin 53706, USA}

\begin{abstract}
The single flux quantum (SFQ) digital superconducting logic family has been proposed for the scalable control of next-generation superconducting qubit arrays. In the initial implementation, SFQ-based gate fidelity was limited by quasiparticle (QP) poisoning induced by the dissipative on-chip SFQ driver circuit. In this work, we introduce a multi-chip module architecture to suppress phonon-mediated QP poisoning. Here, the SFQ elements and qubits are fabricated on separate chips that are joined with In bump bonds. We use interleaved randomized benchmarking to characterize the fidelity of SFQ-based gates, and we demonstrate an error per Clifford gate of 1.2(1)\%, an order-of-magnitude reduction over the gate error achieved in the initial realization of SFQ-based qubit control.  We use purity benchmarking to quantify the contribution of incoherent error at 0.96(2)\%; we attribute this error to photon-mediated QP poisoning mediated by the resonant mm-wave antenna modes of the qubit and SFQ-qubit coupler. We anticipate that a straightforward redesign of the SFQ driver circuit to limit the bandwidth of the SFQ pulses will eliminate this source of infidelity, allowing SFQ-based gates with fidelity approaching theoretical limits, namely 99.9\% for resonant sequences and 99.99\% for more complex pulse sequences involving variable pulse-to-pulse separation.

\end{abstract}

\maketitle

\section{Introduction}

Superconducting qubits have achieved both gate \cite{Barends2014} and measurement \cite{Walter2017, Opremcak2021b} fidelity at the threshold for fault-tolerant operation. \cite{Fowler2012}. Recent demonstrations of quantum supremacy \cite{Arute2019a} and of distance-three and distance-five surface codes \cite{Zhao2021, Krinner2022, Acharya2022} motivate efforts to scale to larger multiqubit arrays that are compatible with robust error correction. However, theoretical estimates suggest that a practical error-corrected quantum computer will require more than one million physical qubits, for physical hardware at the current level of fidelity \cite{Fowler2012}. While it is believed that current quantum--classical interface technology can be scaled by brute force to implement systems with around 1000 physical qubits, limited by the heat load and physical footprint of the control hardware \cite{Krinner2019}, no clear path is known for further scaling up. There have been serious steps to address specific obstacles to scaling, including establishment of quantum links between separated cryogenic systems \cite{Magnard2020}, optimization of room temperature hardware design for a more compact microwave unit \cite{Stefanazzi2022}, integration of cryogenic CMOS-based microwave pulse generators into qubit cryostats for proximal control \cite{Bardin2019, VanDIjk2020}, utilization of low heat-load photonic links to route control and measurement signals within the cryostat \cite{Lecocq2021}, and development of compact, low heat-load microcoax cables \cite{Smith2021}. 
However, an integrated systems engineering approach to scaling superconducting qubits is so far lacking.

\begin{figure}[b!]
\includegraphics[width=\columnwidth]{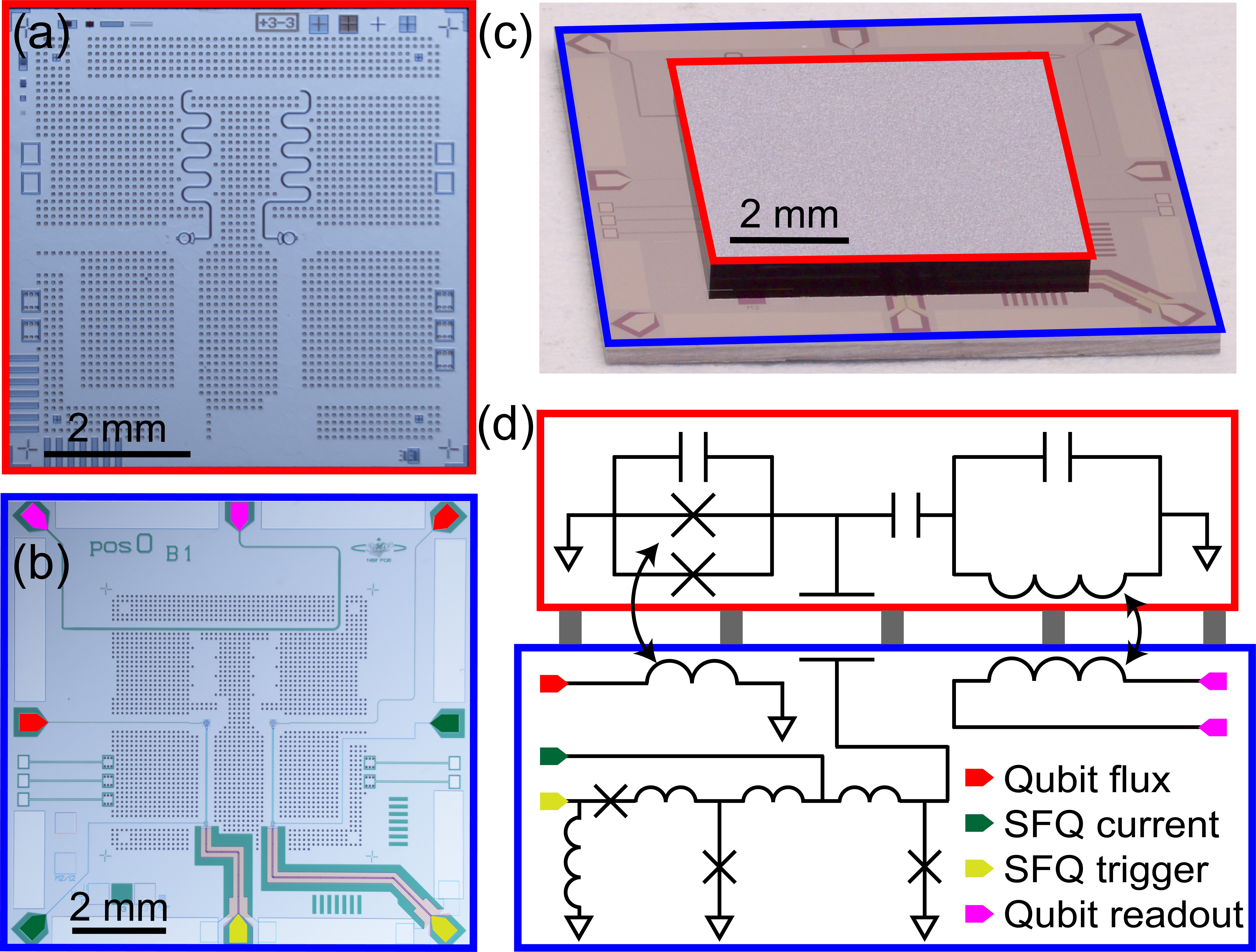}
\caption{Quantum--classical multichip module (MCM).
(a) Micrograph of the qubit chip. Two flux-tunable transmons are fabricated on the chip, each with a local quarter-wave coplanar resonator for readout.
(b) Micrograph of the SFQ driver chip. Two dc/SFQ converters are integrated on the chip, along with the feedline for the readout resonators and flux-bias lines for the qubits. The indium bumps are visible as the regular grid pattern over the continuous ground plane.
(c) Photograph showing the assembled MCM stack; the qubit chip is outlined in red and the SFQ chip is outlined in blue.
(d) Circuit diagram for one qubit--SFQ pair; here, the quarter-wave readout mode is depicted using its lumped-element equivalent. Indium bump bonds between the groundplanes provide the only galvanic connection between the two chips; coupling between circuit elements across the chip-to-chip gap is achieved either capacitively or inductively. The colors used in the legend to identify specific circuit elements are matched to the false coloring of the bond pads in the image of (b).
}
\label{fig:MCM}
\end{figure}

\begin{figure*}[ht!]
\includegraphics[width=\columnwidth]{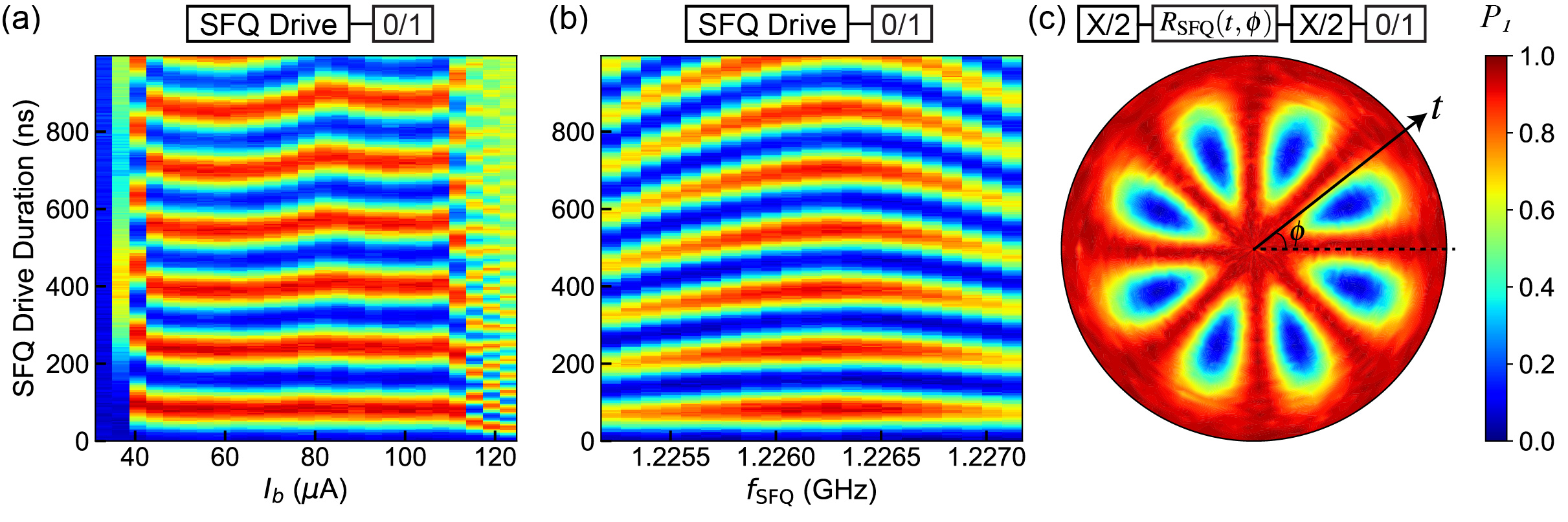}
\caption{SFQ-based qubit operation at subharmonic drive frequency $f_{01}/4$.
(a) Rabi oscillations as a function of current bias $I_{b}$ of the SFQ driver.
(b) Rabi chevron experiment at variable drive frequency in the vicinity of $f_{01}/4$.
(c) Generalized Rabi scan at $f_{01}/4$, with variable time $t$ and phase $\phi$ for the Rabi drive.
}
\label{fig:SFQControl}
\end{figure*}

Typically, qubit control is achieved with shaped microwave pulses derived from room-temperature digital-to-analog converters and microwave generators; the hardware overhead associated with microwave-based qubit control is one of the major obstacles to scaling to large system sizes. An alternative approach is based on the single flux quantum (SFQ) digital logic family \cite{Likharev1991, Mukhanov2011, Soloviev2017}. Here, the qubit mode is irradiated with a train of quantized flux pulses, with pulse-to-pulse timing adjusted to induce a coherent rotation in the computational subspace and to minimize leakage \cite{McDermott2014}. This approach, in conjunction with digital qubit readout using the Josephson photomultiplier \cite{Opremcak2018a, Howington2019, Opremcak2021b}, forms the basis of a scalable quantum--classical interface for ultralarge qubit arrays \cite{McDermott2018}. An initial experiment to implement SFQ-based control of superconducting qubits yielded fidelity of 95$\%$ for $\pi/2$ and $\pi$ rotations \cite{Leonard2019}; gate fidelity was limited by quasiparticle (QP) poisoning \cite{Glazman2021} associated with operation of the dissipative SFQ pulse generator, which was fabricated on the same chip as the qubit. In a more recent experiment, digital control of a 3D transmon qubit with an error per Clifford gate of $2.1(1)\%$ was demonstrated using a Josephson pulse generator \cite{Sirois2021} located at the 3~K stage of the cryostat \cite{Howe2022}. There have been separate theoretical proposals for high-fidelity SFQ-based control sequences involving variable pulse-to-pulse separation \cite{Liebermann2016a, Li2019}, for SFQ-based entangling gates \cite{Jokar2021}, and for a scalable multiqubit architecture based on SFQ control \cite{Jokar2022a}. However, a key prerequisite to adoption of SFQ control for large-scale multiqubit arrays is realization of high-fidelity single-qubit rotations. It is critical to investigate all the potential error channels and to understand the fundamental limits to fidelity, chief among them generation of QPs by the dissipative SFQ pulse driver. 

In this work, we adopt a multi-chip module (MCM) architecture to segregate the SFQ pulse driver and the qubit onto two separate chips. In so doing, we suppress both phonon-mediated QP poisoning and direct diffusion of QPs from the SFQ driver to the qubit. We demonstrate an order-of-magnitude reduction in SFQ-based gate infidelity compared to \cite{Leonard2019}, with an average error per Clifford gate of $1.2(1)\%$. We find that infidelity is dominated by incoherent error associated with \textit{photon}-mediated QP poisoning. We anticipate that straightforward design changes to the SFQ driver and appropriate QP mitigation on the qubit chip can lead to further reductions in gate infidelity, to the point where we are limited by leakage to errors of order 0.1\% with naive resonant control and 0.01\% for optimized control sequences \cite{Liebermann2016a, Li2019}.

This manuscript is organized as follows. In Section \ref{sec:SFQ Control}, we present the quantum--classical MCM and describe basic SFQ-based single-qubit control. In Section \ref{sec:Benchmark}, we describe randomized benchmarking (RB) and interleaved randomized benchmarking (IRB) to characterize the fidelity of $\pi/2$, $\pi$, and average Clifford rotations, and we perform detailed error budgeting of SFQ-based gates. In Section \ref{sec: QP dynamics}, we show that infidelity is dominated by a subtle form of QP poisoning associated with emission of pair-breaking photons from the SFQ driver. In Section \ref{sec: Antenna}, we present simulation results that validate our model for photon-assisted QP poisoning mediated via spurious antenna modes of the qubit and SFQ-qubit coupler. Finally, in Section \ref{sec: Conclusion}, we discuss straightforward modifications to the SFQ-qubit architecture to further suppress QP-induced gate infidelity. These improvements should allow us to access gate fidelity comparable to that achieved with state-of-the-art microwave-based gates, but with a compact, streamlined hardware footprint for the control system.

\section{\label{sec:SFQ Control}Quantum--Classical MCM and SFQ-based Qubit Control}

To suppress QP poisoning, the dominant source of infidelity in previous approaches to SFQ-based qubit control \cite{Leonard2019}, we segregate the qubit and SFQ elements onto two separate chips that are bump-bonded with In to form the MCM stack shown in Fig. \ref{fig:MCM}. The qubit chip shown in Fig. \ref{fig:MCM}(a) incorporates two flux-tunable transmon qubits, each with its own local quarter-wave resonators for state measurement. The SFQ driver chip shown in Fig. \ref{fig:MCM}(b) incorporates two dc-to-SFQ converters along with all control and readout lines for the qubits. The two chips are bonded by In bumps to form the MCM via the technique described in Appendix \ref{App: MCM}; see Fig. \ref{fig:MCM}(c). The MCM architecture suppresses QP poisoning of the qubit in two ways. First, the direct diffusion of QPs from the SFQ driver to the qubit is not possible, as the two elements reside on separate chips that are separated by low-gap In bump bonds: QPs that relax to the In gap edge will be unable to enter the Nb groundplane of the quantum chip \cite{Riwar2019}. Similarly, pair-breaking phonons that propagate to the In bump bonds are expected to scatter to the In gap edge through electron-phonon interaction, so that phonons will have insufficient energy to enter the Nb groundplane of the qubit chip; acoustic mismatch across the bump bonds will further inhibit phonon propagation from chip to chip.  In Fig. \ref{fig:MCM}(d), we show a simplified circuit diagram for one qubit--SFQ pair, with the qubit readout resonator depicted as its lumped element equivalent tank circuit. Each qubit--SFQ pair involves one flux bias line for the qubit, one current bias line for the SFQ driver, and one microwave drive line to trigger SFQ pulses. The qubit readout line is shared between the two qubit--SFQ pairs. Details of the experimental setup are given in Appendix \ref{App: Wiring}.

The quantum--classical MCM is characterized in a closed-cycle dilution refrigerator at a base temperature of 20~mK. Preliminary optimization of SFQ-based qubit rotations is described in Fig.~\ref{fig:SFQControl}. To avoid direct drive of the qubit via crosstalk from the SFQ trigger line, we generate an SFQ pulse train at a subharmonic of the fundamental qubit frequency $f_{01}$. In the experiments described here, we drive the qubit at the frequency $f_{01}/4$. With the SFQ trigger tone applied, we sweep the current bias $I_b$ of the SFQ pulse driver to determine an operating regime where the induced qubit Rabi frequency is insensitive to SFQ driver bias; results are shown in Fig. \ref{fig:SFQControl}(a). Over the optimal range of SFQ driver operation, we find a weak dependence of Rabi frequency on $I_b$; we discuss possible explanations for this dependence in Appendix \ref{App: Stability}. In Fig.~\ref{fig:SFQControl}(b), we show the results of a Rabi chevron experiment used to fine tune the SFQ drive frequency. Finally, we perform the generalized Rabi experiment described in Fig.~\ref{fig:SFQControl}(c) in order to identify the duration and relative timing of SFQ pulse trains needed to execute qubit rotations about arbitrary control vectors oriented in the equatorial plane of the Bloch sphere. 

\begin{figure}[t!]
\includegraphics[width=\columnwidth]{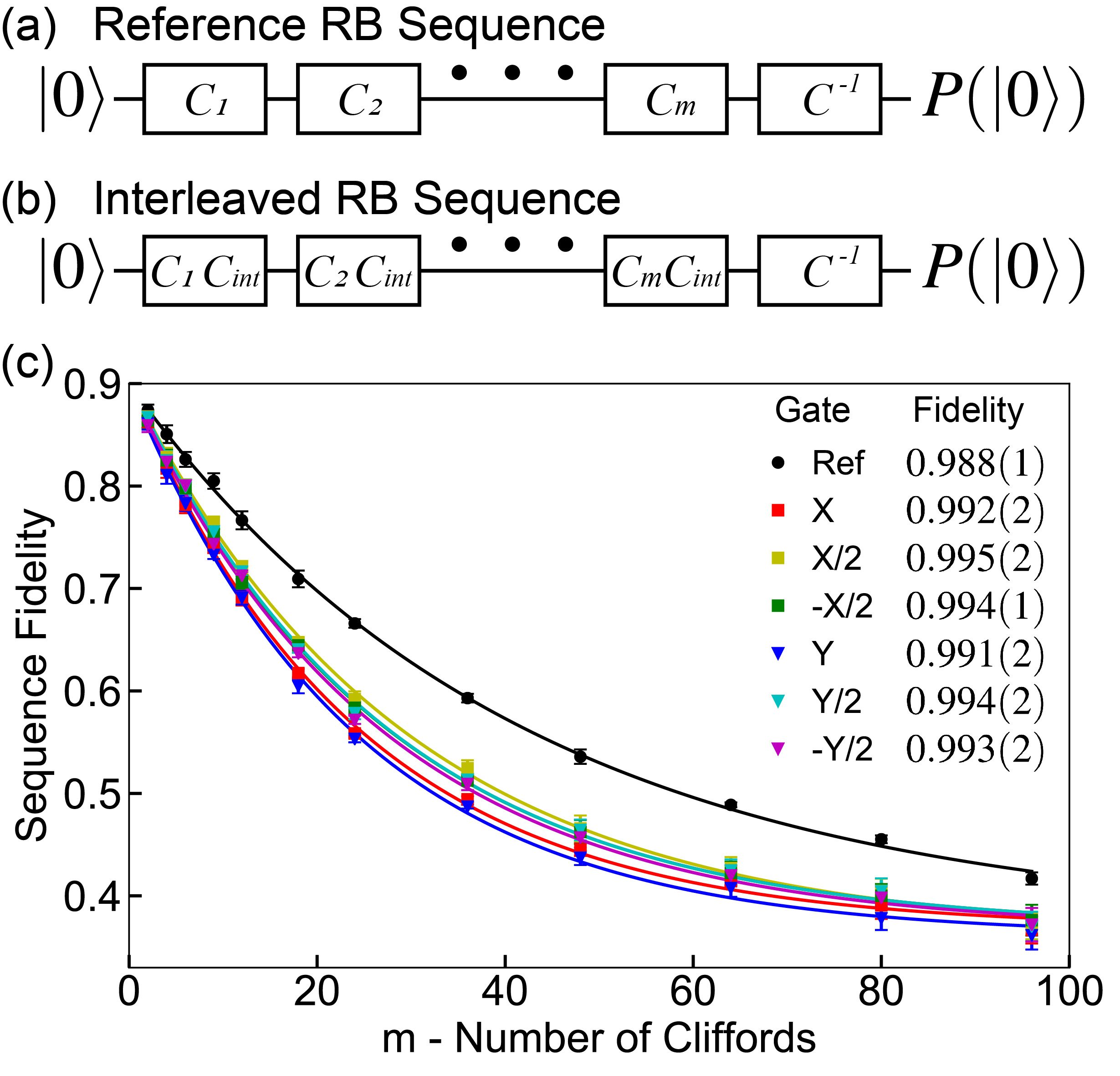}
\caption{Randomized Benchmarking (RB) of SFQ-based gates with drive at frequency $f_{01}/4$~=~1.2264~GHz. Pulse sequences for (a) reference and (b) interleaved RB.
(c) Depolarizing curves for the reference RB sequence and six interleaved RB sequences. Each data point is the average of 150 random sequences across 6 hours. Inset shows the fidelities of the average Clifford gate \cite{Gottesman1997a} and the six interleaved gates.
}
\label{fig:RBIRB}
\end{figure}

\section{\label{sec:Benchmark} Benchmarking of SFQ Gates}

After demonstrating basic qubit control with SFQ pulse trains, we characterize the fidelity of SFQ-based gates. It is critical to find the set of parameters in the multi-dimensional SFQ operation space to optimize gate fidelity. In this work, the SFQ control parameters include the bias current $I_b$, the trigger frequency $f_{\rm SFQ}$, the phase difference between orthogonal rotations, the trigger amplitude, and related mixer calibration parameters of the trigger signal. Here, we follow a two-step optimization procedure involving error amplification to tune up individual gates followed by global optimization using randomized benchmarking (RB) \cite{Knill2008, Chow2010a}. Initially, we find the best operating parameters for each individual gate. We construct equivalent identity and $\pi$-pulse sequences by concatenating many instances of the individual gate, e.g., $S_1= \rm X^{30}$, and $S_2 = \rm X^{31}$. We measure the difference of $\ket{1}$-state occupation following application of the two sequences $P_1(S_2)-P_1(S_1)$ while sweeping the SFQ drive parameters, and we adjust parameters to maximize the sequence contrast. We find that optimal parameters for the SFQ driver are not exactly the same across all SFQ-based gates. For global optimization of SFQ-based gates, including optimization of the time delay associated with orthogonal rotations on the Bloch sphere, we maximize the RB sequence fidelity following the ORBIT method developed in \cite{Kelly2014}. 

Following optimization of SFQ gate parameters, we use the technique of interleaved RB to access gate fidelity independent of state preparation and measurement error \cite{Magesan2012}. Here, we evaluate the fidelity of the set of gates $\{\rm X,  Y, \pm X/2, \pm Y/2\}$. In Fig. \ref{fig:RBIRB}, we present RB data taken across 6 hours, highlighting the temporal stability of SFQ-based single-qubit control. A fit to the reference curve yields average Clifford gate fidelity of $\mathcal{F}_{\rm Cliff} = 0.988(1)$; from the reference gate sequence and the interleaved sequences, we extract the interleaved gate fidelities shown in Fig. \ref{fig:RBIRB}(c). As a check, we calculate the fidelity of an average Clifford gate from the extracted fidelities of the six interleaved gates. The appropriate weighted sum over the interleaved gates fidelities yields an average Clifford gate fidelity of $0.988(3)$, consistent with the measured RB value $\mathcal{F}_{\rm Cliff}$.

\begin{figure}[t!]
\includegraphics[width=\columnwidth]{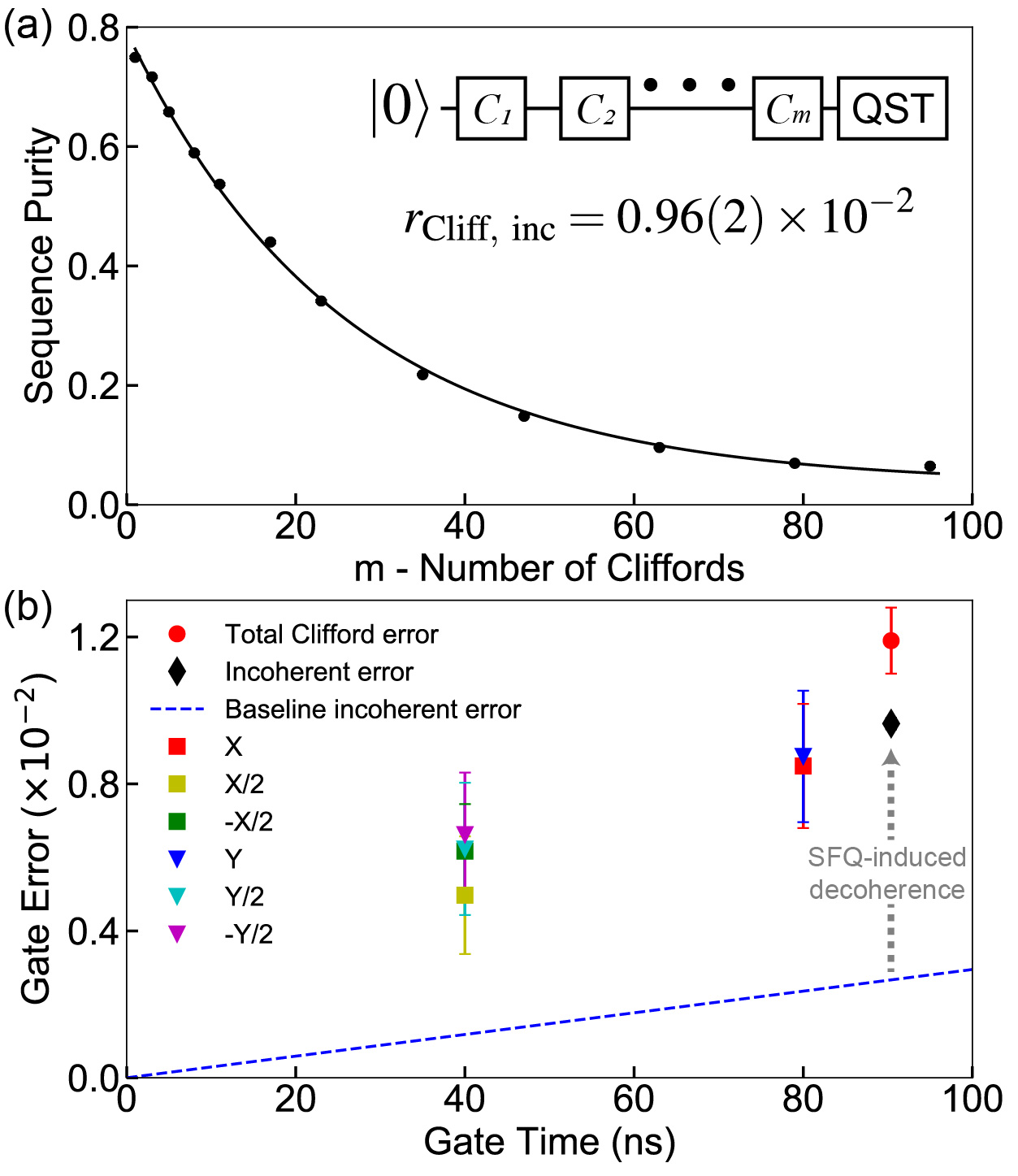}
\caption{
Characterization of incoherent error. (a) Purity benchmarking of SFQ-based gates at pulse frequency $f_{01}/4 =1.2264$~GHz. Inset shows the gate sequence for purity benchmarking. Here, quantum state tomography is applied at the end of the $m$ randomized Cliffords. The incoherent error per Clifford $r_{\rm Cliff, ~inc}$ is extracted from the purity of the final state.
(b) Gate error derived from the RB measurements of Fig.~\ref{fig:RBIRB}(c) versus gate duration. For the average Clifford gate (black diamond), incoherent errors constitute $80\%$ of the total error (red circle). The blue dashed line shows the baseline incoherent error of $0.27\times 10^{-2}$ for the average Clifford gate extracted from the coherence times of the qubit measured with microwave-based gates in the absence of SFQ operation.
}
\label{fig:PB}
\end{figure}

The measured error associated with SFQ-based control shows one order of magnitude reduction compared to the first-generation result of \cite{Leonard2019}. Accurate understanding of the source of the error is critical to further optimization of SFQ-based gates. As a starting point, we need to distinguish between coherent and incoherent error. The former involves pulse sequence miscalibration, while the latter is due to qubit dissipation via uncontrolled coupling to the environment.
Here, we use purity benchmarking \cite{Wallman2015} to extract the incoherent contribution to the error. The gate sequence is shown in Fig. \ref{fig:PB}(a). We apply a random sequence of $m$ Clifford gates followed by quantum state tomography to determine the purity of the qubit state. A fit to the purity curve yields an incoherent error for the average Clifford gate of $r_{\rm Cliff, ~inc} = 0.96(2) \times 10^{-2}$. From RB, we find an error per Clifford gate of $r_{\rm Cliff} = 1-\mathcal{F}_{\rm Cliff} = 1.2(1)\times 10^{-2}$. We see that 80\% of the error is due to incoherent processes. In Fig. \ref{fig:PB}(b), we plot the error of the six interleaved gates characterized previously and of the average Clifford gate versus gate time. The average Clifford gate time is 90.4~ns, 2.26 times that of the $\pi/2$ gate. We see that gate error increases with gate duration. The dashed blue line in Fig. \ref{fig:PB}(b) shows the baseline incoherent error per gate $t_{\rm gate}/T_{\rm error}$ \cite{Barends2015}, where $t_{\rm gate}$ is the gate length and where $1/T_{\rm error} = \frac{1}{3}(1/T_{2, \rm white} + 1/T_1)$ is calculated from the qubit coherence times see Table \ref{tab:qparams} in the appendix for details measured using conventional microwave-based sequences with the SFQ driver turned off \cite{Chen2018}. Comparing this baseline error to the measured incoherent error for the average Clifford gate, we see that operation of the SFQ pulse driver induces additional incoherent error around 2.6 times that of the baseline. In the next section, we show that photon-assisted QP poisoning is the source of the additional incoherent error.

\begin{figure}[b!]
\includegraphics[width=\columnwidth]{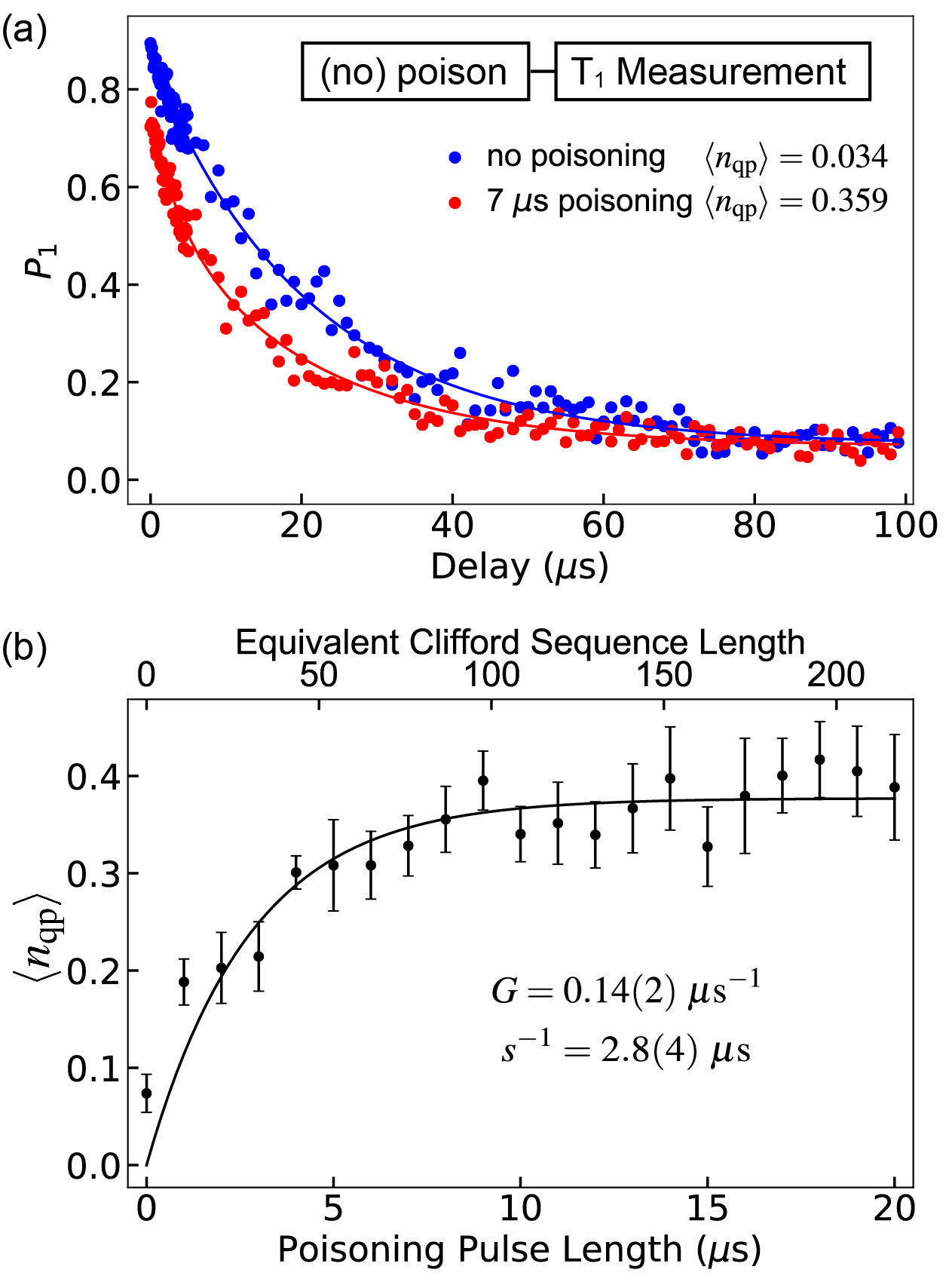}
\caption{
QP poisoning induced by operation of the SFQ pulse driver. 
(a) Microwave-based energy relaxation curves of the qubit with and without prior application of an off-resonant SFQ pulse train at $f_{\rm SFQ}= 1.21$~GHz. The mean number $\langle n_{\rm QP} \rangle$ of QPs coupled to the qubit is extracted from a fit of Eq.~\ref{eq:nonexponential-decay} to the data. 
(b) $\langle n_{\rm QP} \rangle$ vs. SFQ poisoning pulse length. Each point was extracted from eight $T_{1}$ traces of the type shown in (a).
}
\label{fig:QPTrapping}
\end{figure} 

\section{\label{sec: QP dynamics} Dynamics of QP Poisoning}

Nonequilibrium QPs are a dominant decoherence source for superconducting quantum devices \cite{Catelani2012, Serniak2018, Grunhaupt2018, Wilen2021, Siddiqi2021}, and suppression of QP poisoning associated with the dissipative SFQ pulse driver is the primary reason for adopting the MCM architecture described here. The analysis above indicates that SFQ gate infidelity is dominated by incoherent error and that operation of the SFQ pulse driver leads to significant suppression of qubit coherence. QP poisoning remains the most likely mechanism for suppression of qubit coherence. To quantify the generation of QPs at the qubit chip from operation of the SFQ pulse driver, we perform microwave-based inversion recovery experiments following application of an off-resonant SFQ drive, and we fit the recovery signal to the form \cite{Pop2014, Gustavsson2016}
\begin{equation}
    P_1(t)= e^{\langle n_{\rm QP} \rangle \left(\exp (-t/T_{1, \rm qp})-1\right)-t/T_{1, \rm R}},
    \label{eq:nonexponential-decay}
\end{equation}
where $P_1(t)$ is the $\ket{1}$-state occupation of the qubit, $\langle n_{\rm QP} \rangle$ is the mean number of QPs coupled to the qubit, $T_{1, \rm qp}$ is the qubit energy relaxation time per QP, and $T_{1, \rm R}$ is the energy relaxation time due to remaining relaxation channels. In Fig. \ref{fig:QPTrapping}(a), we plot inversion recovery signals measured with (red points) and without (blue points) application of a $7 ~\mu$s off-resonant SFQ poisoning pulse prior to the measurement. The solid traces are fits to the data from Eq.~\ref{eq:nonexponential-decay} to extract $\langle n_{\rm QP} \rangle$. In a separate experiment, we vary the poisoning pulse length prior to the $T_1$ measurement. Fits to the inversion recovery scans yield $T_{1, \rm qp} = 6.8(6) ~\mu \rm s$ and $T_{1, \rm R} = 26(1) ~\mu \rm s$; in Fig. \ref{fig:QPTrapping}(b), we plot $\langle n_{\rm QP} \rangle$ as a function of poisoning length.
The data is well described by a model where the rate of QP removal is linear in QP density, as expected for both diffusion of QPs from the junction and QP trapping at defect sites. We express the time-dependent QP population as follows \cite{Wang2014, Vepsalainen2020a}:
\begin{equation}
    \langle n_{\rm QP} (t)\rangle = \frac{G}{s}(1-e^{-st}),
\end{equation}
where  $G$ is the QP generation rate and $s$ is the QP removal rate. We find $G=0.14(2)~\rm\mu s^{-1}$ and $s^{-1}=2.8(4)~\rm \mu s$. The generation rate corresponds to $3.7(5)\times 10^{-5}$ QPs coupled to the qubit per phase slip of the SFQ driver, a factor of 43 improvement compared to the QP poisoning seen in the first-generation SFQ control experiments of \cite{Leonard2019}.

\begin{figure}[t!]
\includegraphics[width=\columnwidth]{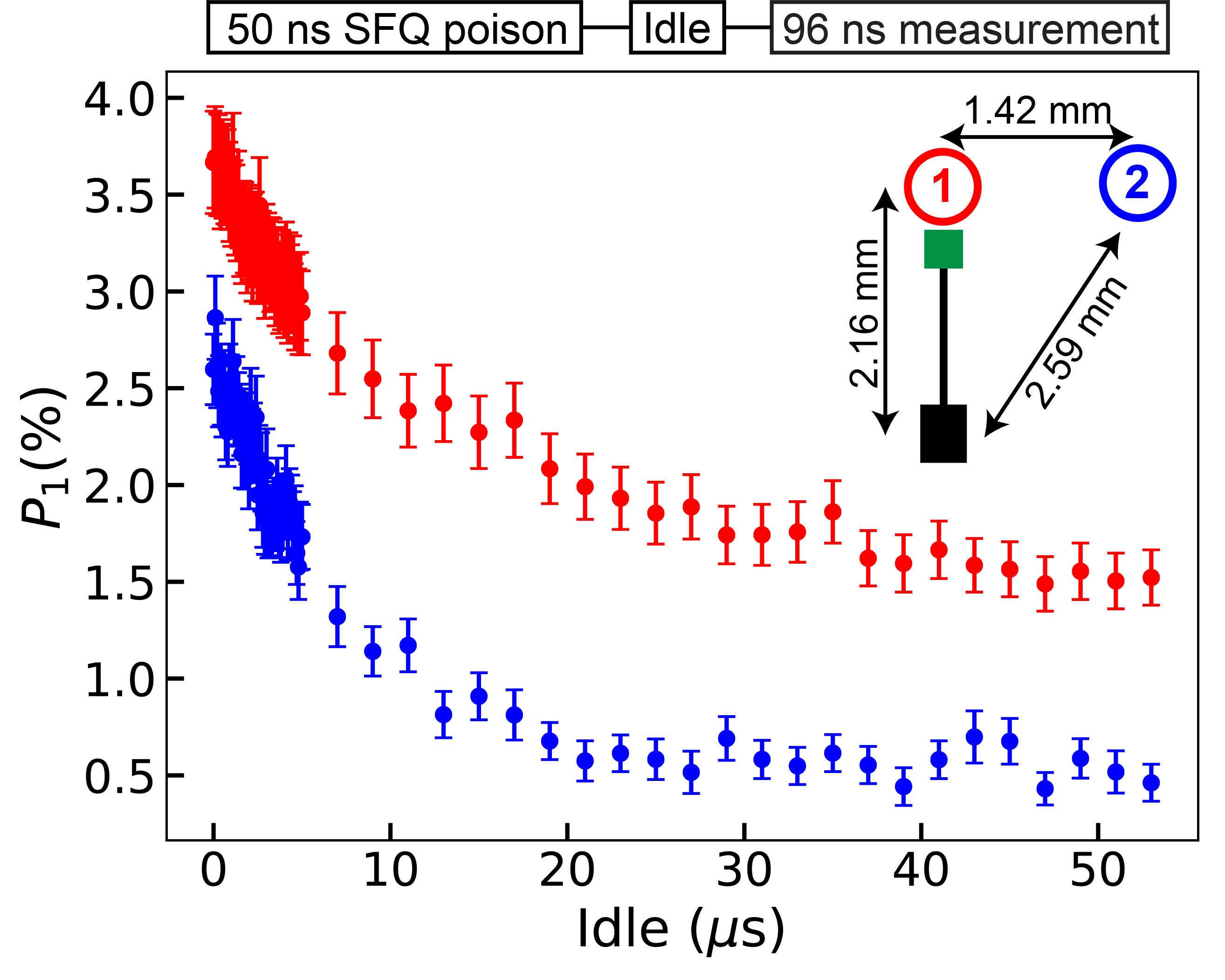}
\caption{Dynamics of QP poisoning from the SFQ pulse driver. A variable idle time follows application of a brief SFQ poisoning pulse; subsequent fast qubit measurement is used to extract the qubit $\ket{1}$-state occupation $P_{1}$. Inset shows the relative positions of $Q_1$ (red), $Q_2$ (blue), the dc/SFQ converter (black rectangle) and the SFQ-qubit coupling capacitor (green square). For both qubits, $P_{1}$ decays monotonically with idle time. We observe no time lag between application of the poisoning pulse and the peaking of enhanced excess $P_1$, indicating that poisoning is mediated by pair-breaking photons, as opposed to pair-breaking phonons.}
\label{fig:PoisonP1}
\end{figure}

Our next task is to understand the physical mechanism for QP poisoning in the MCM architecture. There are two possibilities. In one scenario, QP poisoning is mediated by pair-breaking phonons that propagate from the classical chip to the quantum chip \cite{Patel2017}, despite the presence of the low-gap In bumps that are expected to promote phonon relaxation below the gap edge of the Nb groundplane of the quantum chip. Alternatively, QP poisoning could be dominated by pair-breaking \textit{photons} associated with the ultrahigh-bandwidth SFQ pulses. It is known that qubit structures are efficient absorbers of pair-breaking radiation in the mm-wave range \cite{Rafferty2021b, Liu2022}; for picosecond SFQ pulses with bandwidth of order 100s of gigahertz, the electromagnetic transient could lead to emission of pair-breaking photons that are then absorbed at the qubit junction. We expect to be able to distinguish these two processes by examining the temporal dynamics of QP poisoning in our experiment: while the photon-assisted QP poisoning mechanism will lead to immediate suppression of qubit coherence, the phonon mechanism will involve a time delay of order 10s of $\mu$s between application of the SFQ pulse and the onset of enhanced QP relaxation associated with the propagation of phonons from the SFQ driver to the qubit \cite{Iaia2022c}. 

We probe the dynamics of QP poisoning in the MCM by applying a short 50-ns burst of off-resonant SFQ pulses, using the two qubits on the MCM as QP sensors. We increment the relative delay between application of the poisoning pulse and qubit measurement; in order to access short timescales, we use a fast qubit measurement with duration of 96~ns. The geometry of the experiment is shown in the inset of Fig.~\ref{fig:PoisonP1}. 
We find that the effect of QP poisoning is to increase $P_1$, where the baseline levels of qubits 1 and 2 are 1.6\% and 0.5\% respectively.
This fact alone argues in favor of the photon-mediated mechanism, which drives upward qubit transitions far more efficiently than diffusion across the qubit junction of QPs that are resident in the junction leads \cite{Houzet2019}. We probe qubit population $P_1$ as a function of idle time following the poisoning pulse; results are shown in Fig.~\ref{fig:PoisonP1}. We find a monotonic decrease in excess $P_1$ toward the baseline value following application of the poisoning pulse. In a similar experiment involving injection of pair-breaking phonons into the qubit substrate,  Iaia \textit{et al}. observed a QP-induced enhancement of qubit relaxation rate that peaks at a time $\sim 30 ~\mu$s following application of the poisoning pulse, consistent with diffusive propagation of phonons from the injector junction to the qubit over the $\sim$4-mm separation between the elements \cite{Iaia2022c}.
We take the much faster response of the qubit to the poisoning pulse observed in our experiments as further evidence that coupling of the SFQ driver to the qubit is mediated by photons as opposed to phonons.

\begin{figure}[t!]
\includegraphics[width=\columnwidth]{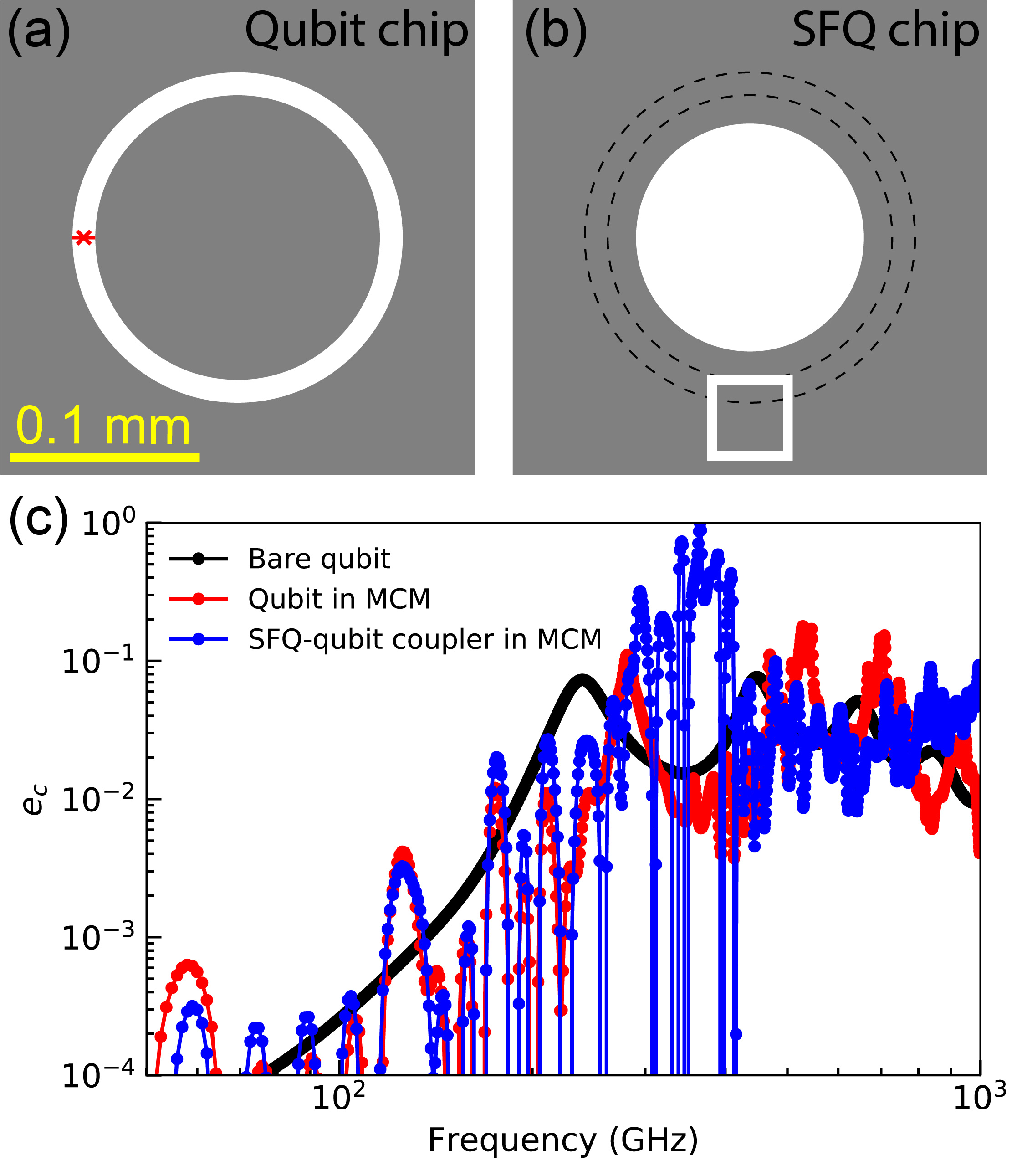}
\caption{Antenna modeling of the qubit and the SFQ-qubit coupler. Schematic of (a) the qubit and (b) the region opposite the qubit on the classical SFQ driver chip. Here, gray represents Nb metallization on the two chips; in the white regions, the Nb has been etched away. The Nb has been removed in a circular region on the classical chip directly opposite the qubit island in order to reduce the capacitance between the qubit and the groundplane of the classical chip. In (a), the position of the qubit junction is indicated by the red cross. In (b), the concentric black dashed lines indicate the position of the gap between the qubit island and groundplane relative to structures on the SFQ driver chip. The patch capacitor that couples SFQ pulses to the qubit is shown at bottom; this SFQ-qubit coupler is fed from the bottom of the image by a microstrip line from the classical SFQ driver circuit. 
(c) Numerically calculated free-space coupling efficiency $e_{c}$ of the antenna modes of the qubit and of the SFQ-qubit coupler \cite{Rafferty2021b}. The black curve represents $e_{c}$ of the qubit with the SFQ chip removed; the dominant feature at 250~GHz corresponds to the fundamental full-wave resonance of the aperture antenna formed by the qubit island embedded in the groundplane. The red and blue traces are simulation results for $e_{c}$ of the qubit and SFQ coupler modes calculated in the full MCM architecture. 
}
\label{fig:AntennaEc}
\end{figure}

\section{\label{sec: Antenna} Antenna Coupling of the SFQ Transient to the Qubit}

It has recently been shown that absorption of pair-breaking photons is a dominant source of QP poisoning in Josephson devices \cite{Houzet2019, Pan2022a, Liu2022, Diamond2022}. For typical geometries, the superconducting qubit structure forms a resonant antenna that provides an efficient power match from free space to the high-impedance Josephson junction at mm-wave frequencies \cite{Rafferty2021b}. Due to its short temporal duration of order ps, the bandwidth of a single SFQ pulse is of order 100s of gigahertz, sufficient to excite the mm-wave antenna mode of the qubit and generate QPs. In Fig.~\ref{fig:AntennaEc}, we consider the antenna modes of the qubit and SFQ coupler structures used in our experiments. The experimental geometry is shown in Fig.~\ref{fig:AntennaEc} (a-b); here, Nb metallization on the quantum and classical chips is shown in gray, and the white regions indicate where the Nb has been removed. The circular qubit and the SFQ-qubit capacitive coupler both act as resonant aperture antennas with efficient coupling to free space at mm-wave frequencies. Following the modeling described in \cite{Rafferty2021b}, we plot the free-space coupling efficiency $e_c$ of the bare qubit antenna mode as the black curve in Fig. \ref{fig:AntennaEc}(c). The structure shows a clear resonance corresponding to a match between the qubit perimeter and one full wavelength of the radiated field. In the same figure, we plot the coupling efficiency of the qubit (red curve) and the SFQ coupler (blue curve) calculated for the full MCM structure. Both modes display a complicated frequency-dependent coupling efficiency with peaks in the resonant response around 300~GHz. Similarity in the coupling efficiency of the qubit and SFQ coupler could be due to loading of both modes by the etched cavity in the groundplane of the SFQ driver chip; see Appendix \ref{App: MCM}.

After verifying that the SFQ-qubit pair can be considered as a coupled antenna system, we show that the ps-scale SFQ pulse indeed leads to emission of photons with energy sufficient to break Cooper pairs; see Appendix \ref{App: Gaussian SFQ} for a detailed discussion.

From our analysis of QP poisoning presented in Fig.~\ref{fig:QPTrapping}, a single SFQ pulse delivered to the qubit generates $1.1\times10^{-4}$~QP at the qubit junction. We can ask the question: if all the energy that goes into QP generation is derived from SFQ-qubit antenna coupling, what is the efficiency of energy transfer from the driver to the qubit? We define an energy conversion factor $\alpha$ as follows: 
\begin{equation}
    \alpha E_{\rm SFQ} = 1.1\times10^{-4}\times E_{\rm qp} ,
\end{equation}
where $E_{\rm SFQ}$ is the available energy from one SFQ pulse and $E_{\rm qp}$ is the energy of one QP. We take the available energy per SFQ pulse to be equal to the phase slip energy, and for simplicity we assume that the generated QPs are concentrated at the gap edge of Al. We find
\begin{align}
    E_{\rm SFQ} &  = I_c \Phi_0, \\
    E_{\rm qp} &= h\times 50 ~\rm GHz,
\end{align}
where $I_c \sim$~100~$\mu$A is the critical current of the output junction of the SFQ driver. 
Solving the equations above, we find an experimental energy conversion efficiency $\alpha_{\rm exp} =1.8\times 10^{-8}$. 

\begin{figure}[t!]
\includegraphics[width=\columnwidth]{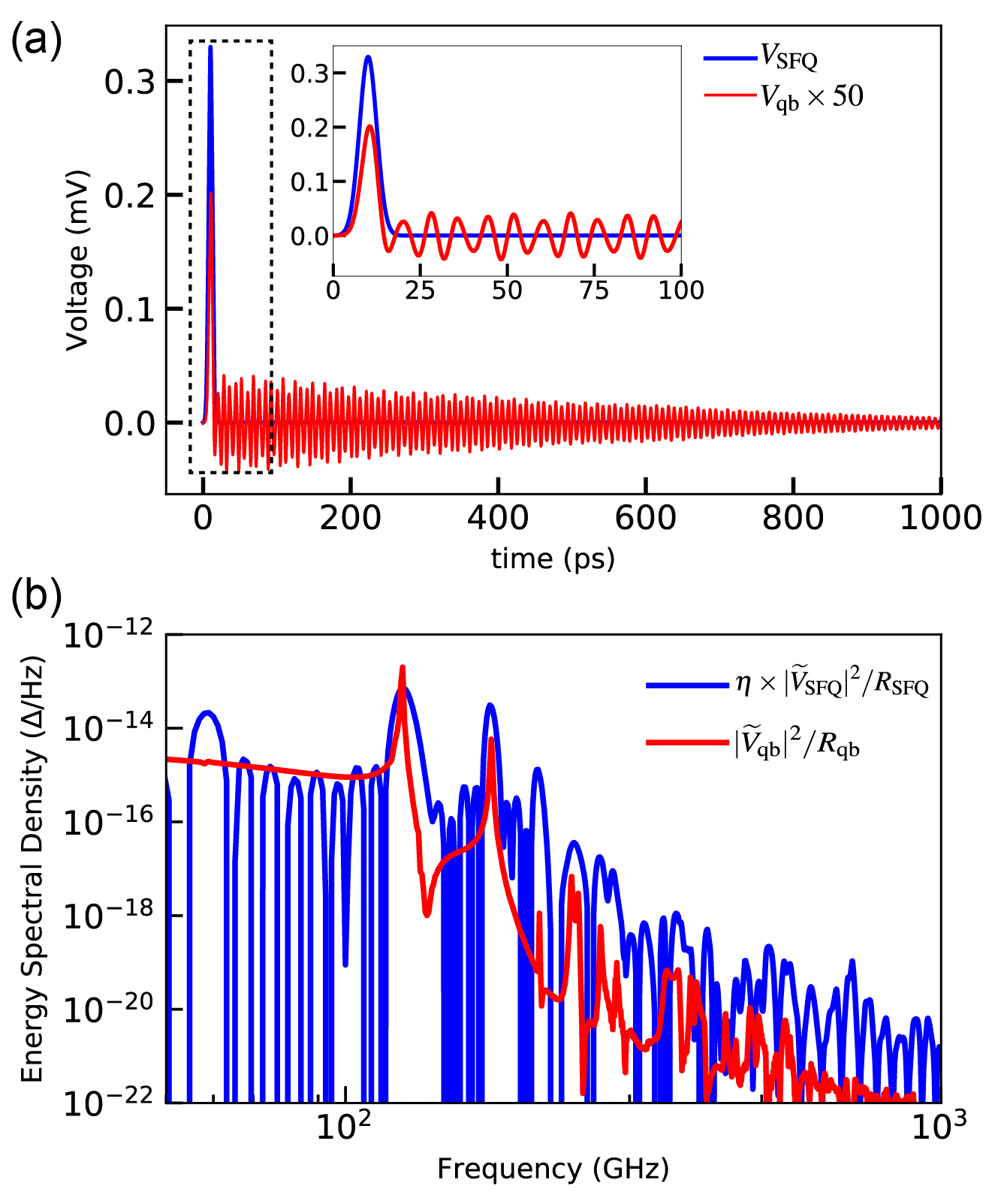}
\caption{
Temporal response of the qubit antenna mode to a single SFQ pulse.
(a) SFQ pulse and induced transient at the qubit. A single SFQ pulse $V_{\rm SFQ}$ with $\sigma=2.5$~ps is delivered from the SFQ-qubit coupler (blue trace). Antenna coupling between the SFQ-qubit coupler and the qubit results in an induced voltage $V_{\rm qb}$ at the qubit port (red trace; voltage scaled by a factor of 50 for clarity). 
(b) Energy spectral density of the qubit response expressed in units of $\Delta/ \rm Hz$, where $\Delta = E_{\rm qp} = h\times 50 ~\rm GHz$. The red trace is the energy spectral density calculated from the qubit transient in (a); here, $\widetilde{V}_{\rm qb}$ is the Fourier transform of the induced voltage at the qubit, and $R_{\rm qb}$ is the normal state resistance of the qubit junction. The blue trace shows the product of the photon coupling efficiency $\eta$ based on the frequency-domain antenna modeling and the energy spectral density of the SFQ pulse. The dominant Fourier component in the qubit response matches the product of the energy spectral density of the SFQ pulse and the photon coupling efficiency of the coupler-qubit system. 
}
\label{fig:AntennaTime}
\end{figure}
We can compare the experimentally extracted $\alpha$ with the result derived from time-domain modeling of the coupled antenna system. We take $R_{\rm SFQ} = 1 ~\Omega$ as the source impedance associated with the SFQ-qubit coupler, and we model the SFQ transient as a Gaussian pulse with width $\sigma = 2.5$ ps, derived from WRspice \cite{WRSpice} simulation of our dc/SFQ converter. As shown in Fig.~\ref{fig:AntennaTime}(a), the SFQ pulse applied to the coupler (blue) induces an oscillatory voltage response at the qubit junction port (red, scaled by a factor 50), with a dominant frequency set by the product of the energy spectral density of the SFQ pulse and the photon coupling efficiency of the coupler-qubit system. In Fig.~\ref{fig:AntennaTime}(b), we show in red the energy spectral density of the response signal at the qubit port $|\widetilde{V}_{\rm qb}|^2/R_{\rm qb}$; here, $\widetilde{V}_{\rm qb}$ is the Fourier transform of the voltage induced at the qubit port and $R_{\rm qb} = 8.0~\rm k\Omega$ is the normal state resistance of the junction. As in \cite{Liu2022}, we define the photon coupling efficiency $\eta$ as:
\begin{equation}
    \eta(f) = e_{c, \rm qb} \, e_{c, \rm coupler},
\end{equation}
where $f$ is frequency and $e_{c, \rm qb} \, (e_{c, \rm coupler})$ is the coupling efficiency of the qubit (coupler) calculated in Fig.~\ref{fig:AntennaEc}(c). In Fig.~\ref{fig:AntennaTime}(b), we plot in blue the available energy spectral density from the SFQ pulse $| \widetilde{V}_{\rm SFQ}|^2/R_{\rm SFQ}$ scaled by the coupling efficiency $\eta$. The close agreement with the spectrum calculated from the voltage transient at the qubit junction suggests that transport of pair-breaking energy between the SFQ coupler and the qubit is dominated by direct antenna coupling between the structures. 

From our modeling, we calculate the total energy dissipated at the qubit port for photons with a frequency above 100~GHz, and we find a simulated energy conversion factor $\alpha_{\rm sim}(\sigma=2.5\rm ~ps)=3.8\times 10^{-8}$, in reasonable agreement with the experimental value $\alpha_{\rm exp} =1.8\times 10^{-8}$. For the sake of completeness, we have also simulated antenna coupling to the qubit of SFQ pulses with varying widths. For SFQ pulse widths $\sigma=0.5, 1, 2$, and $5\rm ~ps$, we find energy conversion efficiency $\alpha_{\rm sim}=1.3\times 10^{-5}, 1.3\times 10^{-6}, 9.7\times 10^{-8}$, and $1.0\times 10^{-12}$, respectively. It is clear that broader SFQ pulses, for which the energy is compressed into a narrower spectral band, lead to a suppression of antenna-mediated QP poisoning at the qubit. As the typical qubit oscillation period $\sim$200~ps is orders of magnitude longer than typical SFQ pulse widths, a straightforward redesign of the SFQ driver to suppress the spectral weight of the SFQ transient above 100~GHz provides an obvious path to eliminating photon-mediated QP poisoning. Deviation of the SFQ pulse from the ideal delta-function will cause misrotation of the qubit state vector on the Bloch sphere \cite{Howe2022}; however, this is a coherent error that is readily addressed by appropriate gate calibration.

\section{\label{sec: Conclusion} Conclusion}

In this work, we have advanced the state of the art for SFQ-based digital control of superconducting qubits. By segregating qubits and classical control elements on separate chips in an MCM architecture, we suppress phonon-mediated QP poisoning and achieve an error per Clifford gate of 1.2(1)\%. This gate infidelity represents a one-order-of-magnitude reduction compared to the first demonstration of SFQ-based qubit control \cite{Leonard2019}, and a factor-of-two reduction in the infidelity achieved in recent work involving a 3D transmon controlled by a Josephson pulse generator located at the 3~K stage of the cryostat \cite{Howe2022}. We find that residual gate infidelity is dominated by photon-assisted QP poisoning mediated via spurious mm-wave antenna modes of the qubit and SFQ-qubit coupler. To suppress QP generation at the qubit from the high-bandwidth SFQ pulse, we suggest a modest redesign of the SFQ driver circuit to yield SFQ pulses with broader characteristic temporal width, corresponding to a narrower pulse bandwidth in the frequency domain. Such a redesign will concentrate the power emitted by the SFQ driver below the aluminum gap edge, so that QP generation is not possible. The qubit and the SFQ-qubit coupler could also be modified to suppress their antenna coupling to free space at frequencies just above the aluminum gap \cite{Rafferty2021b, Liu2022, Pan2022a}. To protect the qubit from any residual nonequilibrium QPs, appropriate superconductor gap engineering \cite{Riwar2019, Kalashnikov2020, Martinis2021a, Bargerbos2022, Catelani2022} could be harnessed to promote the rapid outflow of QPs from the qubit junction and to prevent the inflow to the junction of QPs from remote parts of the qubit circuit. With these steps to mitigate the various forms of nonequilibium QP poisoning, SFQ gate fidelity of $99.9\%$ is achievable using resonant SFQ pulse trains \cite{McDermott2014}. Ultimately, more complex control sequences involving nonuniform SFQ pulse spacing should enable single-qubit gate fidelity of  $99.99\%$, on par with that achieved using microwave-based gates, but with a significant reduction in hardware footprint for the control system.

\section{Acknowledgments}
We thank E. Leonard Jr. and M. A. Beck for stimulating discussions. C. H. L. was additionally funded by NSF award DMR-1747426. We thank M. Castellanos-Beltran and A. Sirois of NIST for assistance with SFQ-driver design, layout and testing. This research was sponsored in part by the WARF Accelerator. SFQ-driver and fabrication work was partially supported by the Office of the Director of National Intelligence (ODNI), Intelligence Advanced Research Projects Activity (IARPA), under Interagency Agreement IARPA-20001-D2022-2203120004. Portions of this work were performed at the UW-Madison Wisconsin Centers for Nanoscale Technology, partially supported by the NSF through the University of Wisconsin Materials Research Science and Engineering Center (DMR-1720415). This This work was performed in part under the auspices of the U.S. Department of Energy by Lawrence Livermore National Laboratory under Contract No. DE-AC52-07NA27344. We gratefully acknowledge support from the NIST Program on Scalable Superconducting Computing and the National Nuclear Security Administration Advanced Simulation and Computing Beyond Moore's Law program (Grant No. LLNL-ABS-795437)

\clearpage

\appendix

\renewcommand{\thefigure}{S\arabic{figure}}
\setcounter{figure}{0}
\renewcommand{\thetable}{S\arabic{table}}
\setcounter{table}{0}

\section{Fabrication of the Quantum-Classical MCM}

\subsection{Qubit}

The qubit chip was fabricated on a high resistivity ($>10~\rm k\Omega$-cm) 3-inch Si substrate with 100 crystal orientation. The native oxide of the silicon is stripped in dilute ($2\%$) hydrofluoric acid for 60 seconds immediately prior to transfer of the wafer into the sputter deposition chamber used for growth of the 100-nm Nb base electrode. We use an i-line projection lithography tool to define the qubit islands and readout resonators. We then etch the Nb using $\rm BCl_3/Cl_2$ chemistry in an inductively coupled plasma reactive ion etch tool. We use an electron-beam writer to define the Dolan bridges for the qubit junctions; the $\rm Al/AlO_x/Al$ junctions are then formed by double-angle evaporation and thermal oxidation in an electron-beam evaporator.

\subsection{SFQ Driver}

The driver circuits were fabricated at the NIST Boulder Microfabrication Facility. The substrates used were 3--inch Si wafers with 150 nm of thermal oxide. The fabrication of the drivers was based on the process for SFQ digital circuits described in \cite{Olaya2019} with the following changes: 1) only three superconducting layers were used, requiring only one chemical--mechanical--planarization step before deposition of the second Nb layer; 2) the top Nb layer was used as the ground plane; 3) the barrier material used for the SIS junctions was amorphous Si, and external shunt resistors were needed to bring the damping of these SIS junctions near the critical regime with Stewart-McCumber parameter $\beta_{c}~\sim~2$; 4) the critical current density of the junctions is 1 kA/cm$^2$ and the characteristic junction frequency is 240~GHz; and 5) the shunt resistors were made of palladium/gold alloy films with sheet resistance of 2 $\Omega$/$\square$ deposited by electron--beam evaporation and defined by a two--layer lift--off resist process.

\subsection{\label{App: MCM}MCM}

The fabrication steps for the under-bump and indium bump layers deposited on each chip are similar to those described in \cite{Lucas2022} but with a target bump thickness of 5~$\mu$m. Additionally, hydrogen plasma cleaning before bonding is avoided to prevent potential contamination of Nb layers used both in the SFQ and qubit chips. The MCM is bonded in a commercial flip-chip bonder. Before bonding, the components are aligned to $\pm~1.0~\mu$m. The coplanarity is adjusted to be less than $100~\mu$radians. The bonding force used was 21.6~kN, which, given optical profilometer measurements of the bump thickness ($4.6~\pm~0.2~\mu$m) and top contact area (260~$\mu$m$^2$) for the 2193 bumps on each chip, results in a calculated effective pressure of 3.9$\times10^{10}$~N/m$^2$ and an expected resulting chip gap of $6.4~\pm~0.3~\mu$m.

\section{\label{App: Wiring}Wiring}

\begin{figure*}[t]
\includegraphics[width=\textwidth]{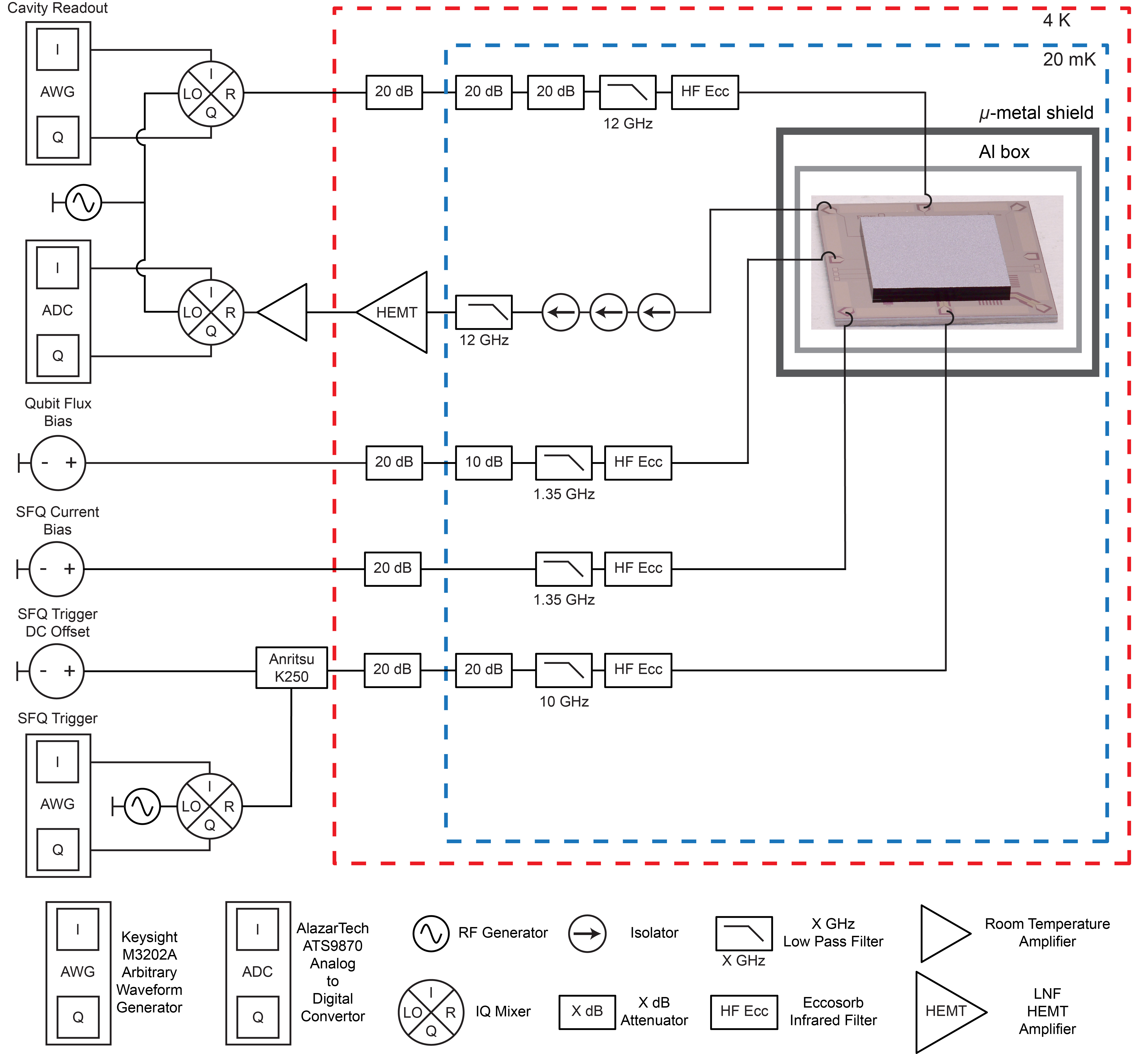}
\caption{\textbf{Wiring diagram of the experiments.}}
\label{fig:wiring_diagram} 
\end{figure*}

The experimental setup and wiring are shown in Fig. \ref{fig:wiring_diagram}.

\section{\label{App: Stability} Impact of SFQ pulse errors}

\begin{figure}[t!]
\includegraphics[width=\columnwidth]{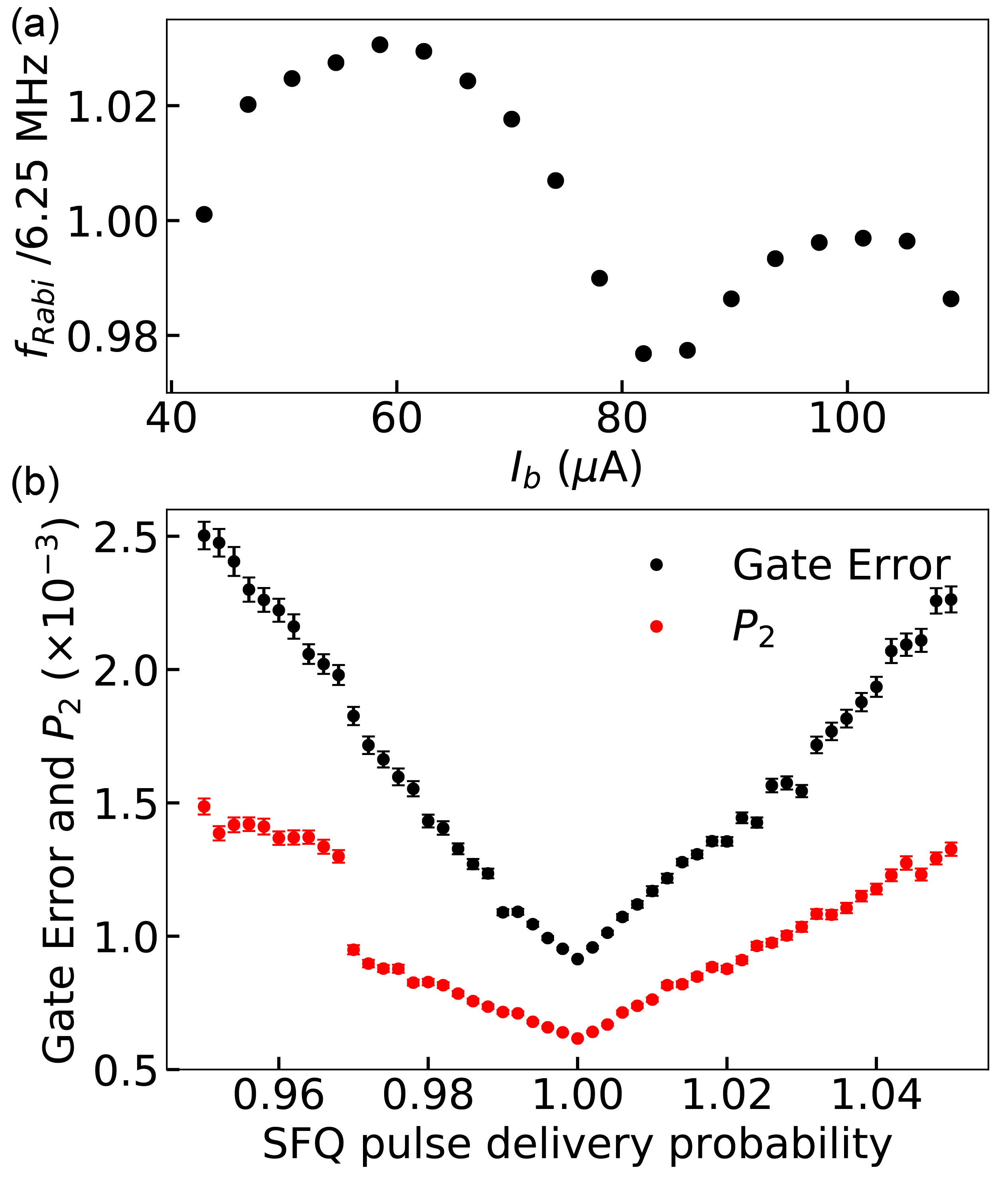}
\caption{
Stability of SFQ pulse delivery and its effect on qubit control.
(a) Scaled Rabi oscillation frequency $f_{\rm Rabi}$ extracted from Fig. \ref{fig:SFQControl}(a) as a function of $I_b$. Over a bias current range from 40 to 120 $\mu$A, we find approximately $\pm$ 3\% variation in $f_{\rm Rabi}$ around the value 6.25~MHz. A possible explanation for the variation in Rabi frequency is that the number of SFQ pulses delivered per cycle of the trigger waveform is either less than or greater than one.
(b) Simulated gate error and leakage to the $\ket{2}$ state for a $\pi/2$ gate implemented with an imperfect SFQ pulse driver. Gate error and leakage both increase in the presence of SFQ pulse dropouts and double pulses.
}
\label{fig:UnsteadySFQ}
\end{figure}

In addition to incoherent error from QP poisoning, 
the instability of SFQ pulse delivery represents another potential error channel. In Fig.\ref{fig:SFQControl}(a), we see that the Rabi frequency associated with resonant SFQ drive is not a constant over the full range of current bias; similar behavior was seen in \cite{Leonard2019, Howe2022}. In Fig.~\ref{fig:UnsteadySFQ}(a), we plot the extracted Rabi frequency versus bias current.  We find relative variation in the Rabi frequency of order 5\% over the bias current range considered. In the following, we examine the possibility that this variation is due to SFQ driver errors, namely, missed SFQ pulses or delivery of double pulses, and we estimate the resulting contribution to SFQ gate error. Pulse dropouts will result in systematic underrotation of the qubit state and a reduction in Rabi frequency, while double pulses will cause overrotation of the qubit and an increase in Rabi frequency.

We define the SFQ pulse probability $P_{\rm SFQ}$ in three regimes. For $P_{\rm SFQ}<1$, one pulse is delivered per clock cycle with probability $P_{\rm SFQ}$; the probability of a pulse dropout is $1-P_{\rm SFQ}$. For $P_{\rm SFQ}=1$, exactly one SFQ pulse is delivered per cycle of the trigger waveform. Finally, for $P_{\rm SFQ}>1$, a double SFQ pulse is delivered with probability $P_{\rm SFQ}-1$, while a single SFQ pulse is delivered with probability $2-P_{\rm SFQ}$.

Following \cite{McDermott2014}, we perform Monte Carlo simulations of the gate error and leakage to the $\ket{2}$ state for a $Y/2$ gate realized with the parameters of the SFQ driver-qubit pair used in these experiments; simulation results are shown in Fig.~\ref{fig:UnsteadySFQ}(b). We find increased gate error and leakage as the probability of pulse dropouts or double pulses increases. For pulse dropout or double pulse probability less than 3\%, compatible with the variation in Rabi frequency over the bias current range from 40~$\mu$A to 120~$\mu$A, we put an upper bound on gate error of $\sim~0.2\%$. While instability of the SFQ driver does not currently limit gate fidelity, it is possible that SFQ pulse errors will be a dominant source of infidelity in SFQ gates once QP poisoning is fully suppressed.

\begin{figure}[b!]
\includegraphics[width=\columnwidth]{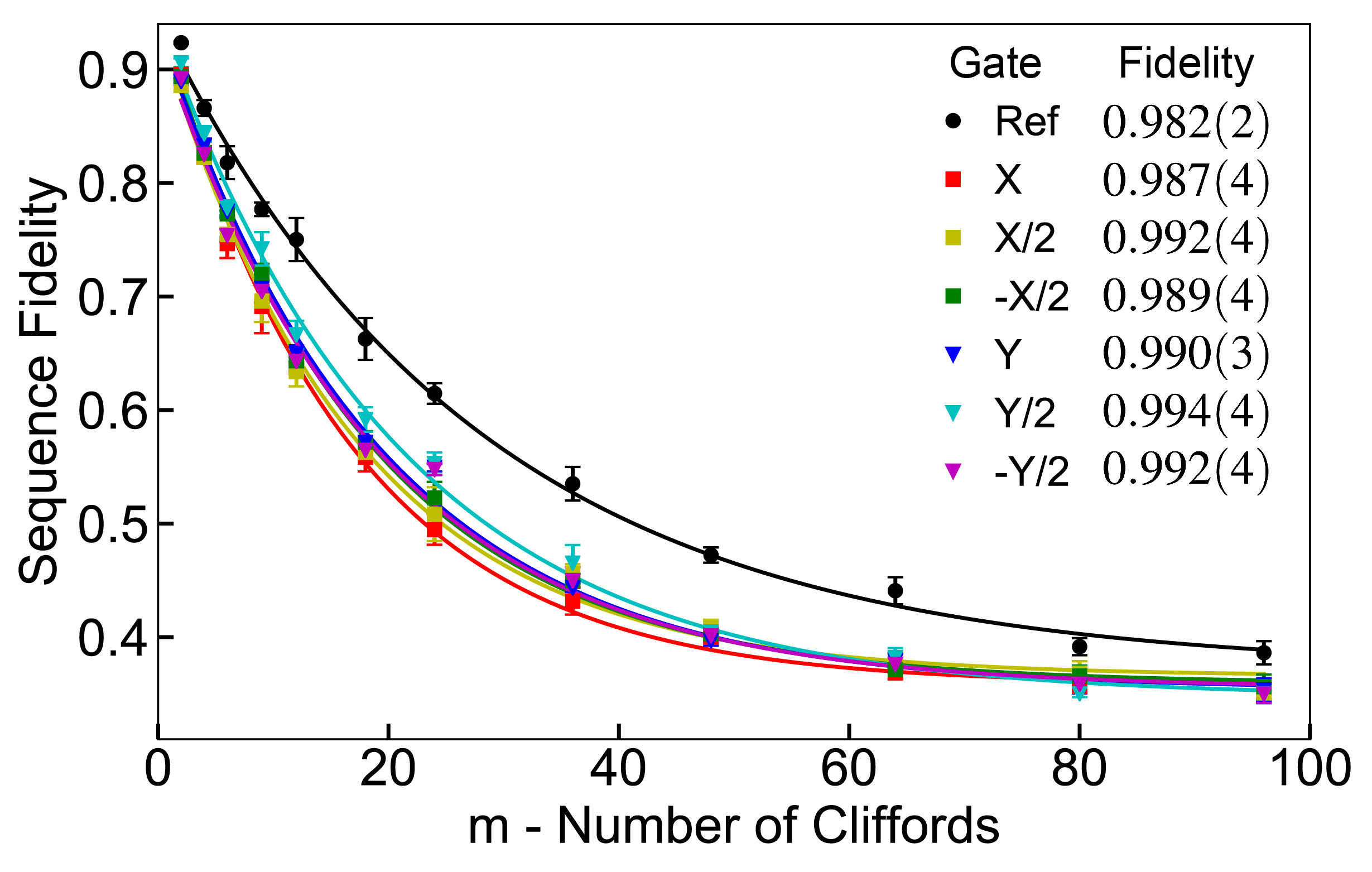}
\caption{IRB of resonant SFQ-based gates implemented at drive frequency $f_{01}/2$.
}
\label{fig:RBIRBOver2}
\end{figure}

\section{SFQ-qubit Parameters}

In Table \ref{tab:qparams}, we list the measured and extracted parameters of the SFQ-qubit pair used in these experiments.

\begin{table}[h]

\caption{\textbf{Parameters of devices used in the experiments.} Resonator and qubit frequencies are measured by spectroscopy. Qubit baseline coherence times $T_1$ and $T_{2, \rm white}$ are extracted from microwave-based inversion recovery and echo sequences. $T_{1,\rm qp}$ is extracted from nonlinear fits to the inversion recovery signals shown in Fig.~\ref{fig:QPTrapping}. The SFQ operation parameters, including trigger frequency, trigger power, trigger DC offset and current bias, are chosen to maximize SFQ-based qubit gate fidelity. The SFQ-qubit coupling capacitance is calculated following \cite{McDermott2014} from the measured Rabi frequency shown in Fig.~\ref{fig:UnsteadySFQ}(a). 
}

\begin{tabular}{ |l|c|c| }
\hline
Description & Symbol & Value \\
\hline
\hline
Readout resonator frequency & $f_{\rm RO}$  & 6.786 GHz\\
\hline
Qubit max operating frequency & $f_{01}$  & 4.906 GHz\\
\hline
Qubit energy relaxation time & $T_{1}$  & 26 $\mu$s\\
\hline
Qubit phase relaxation time & $T_{2, \rm white}$  & 20 $\mu$s\\
\hline
Qubit energy relaxation time per QP & $T_{1, \rm qp}$  & 6.8 $\mu$s\\
\hline
SFQ trigger frequency & $\omega _{\rm SFQ}/2\pi$  & 1.226 GHz\\
\hline
SFQ trigger power &   &  -45 dBm \\
\hline
SFQ trigger DC offset &   & 90 $\mu$A\\
\hline
SFQ current bias & $I_b$  & 80 $\mu$A\\
\hline
SFQ-qubit coupling capacitance & $C_{\rm SFQ}$  & 180 aF\\
\hline
\end{tabular}
\label{tab:qparams}
\end{table}

\section{Characterization of SFQ-based gates with drive at $f_{01}/2$}

We have also used IRB to characterize SFQ-based control at the first subharmonic $f_{01}/2$ of the qubit fundamental frequency; results are shown in Fig. \ref{fig:RBIRBOver2}. The average error per Clifford gate is 1.8(2)\%, which is slightly higher than the result obtained at a drive frequency $f_{01}/4$. It is likely that the degraded fidelity at the higher drive frequency is due to the higher rate of photon-assisted QP generation associated with the shorter interpulse spacing.

\section{\label{App: Gaussian SFQ}Frequency-domain analysis of SFQ pulses}

\begin{figure}[t]
\includegraphics[width=\columnwidth]{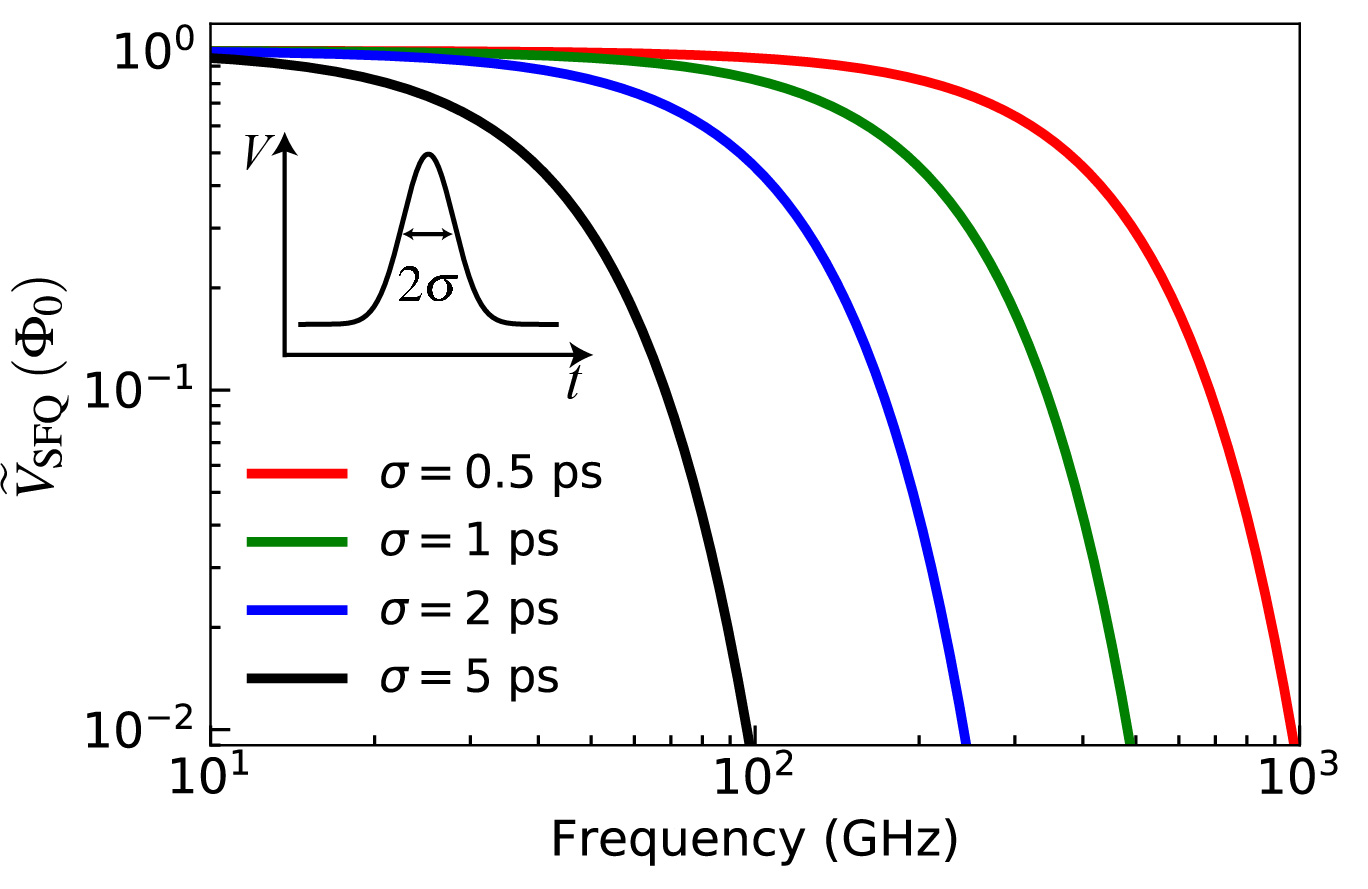}
\caption{Fourier transform of Gaussian SFQ pulses with varying standard deviation $\sigma$ from 0.5-5~ps. In the Fourier domain representation of the pulse, the standard deviation is $\sigma_{f} = (2\pi\sigma)^{-1}$. Narrower SFQ pulses in the time domain involve more spectral weight above the Al energy gap $2\Delta_{\rm Al}/h \sim 100$~GHz, where there is the possibility of photon-assisted QP generation. 
}
\label{fig:SFQGaussian}
\end{figure}

We consider an SFQ pulse with Gaussian envelope in the time domain:
\begin{equation}
    V_{\rm SFQ}(t) = \frac{\Phi_0}{\sqrt{2\pi}\sigma}e^{\frac{-t^2}{2\sigma^2}},
\end{equation}
where $\Phi_0$ is the magnetic flux quantum and $\sigma$ is the standard deviation of the pulse in time. The Fourier transform of the SFQ pulse is given by
\begin{align}
    \widetilde{V}_{\rm SFQ}(f) &=\Phi_0 e^{-\frac{f^2}{2\sigma_f ^2}}, 
\end{align}
where $\sigma_{f} = (2\pi\sigma)^{-1}$ is the standard deviation of the pulse in the frequency domain. For typical Nb-based SFQ devices \cite{Likharev1991}, $\sigma$ is around 1~ps. In Fig.~\ref{fig:SFQGaussian}, we plot $\widetilde{V}_{\rm SFQ}(f)$ for Gaussian SFQ pulses with four values of $\sigma$. For shorter pulses, the SFQ transient involves significant spectral weight at frequencies above the Al energy gap ($\sim$ 100 GHz). Pair-breaking photons emitted by the transient can couple resonantly to the qubit structure via the spurious mm-wave antenna modes of the device. To suppress this form of QP poisoning, it is possible to intentionally broaden the SFQ pulses by increasing the damping of the SFQ driver. As the qubit oscillation period is two orders of magnitude larger than the SFQ pulse width, broadening the SFQ pulse to, say, $\sigma$~=~5~ps will have negligible effect on the coherent qubit rotation induced by the SFQ pulse. 

\clearpage
\bibliography{references, local}

\end{document}